\documentclass{iau}

\usepackage{amsmath}
\usepackage{graphicx}

\usepackage{subcaption} 
\usepackage{multirow} 

\usepackage{natbib} 
\usepackage{aas_macros} 
\bibpunct{(}{)}{;}{a}{}{,} 

\usepackage{hyperref} 
\hypersetup{colorlinks=true, linkcolor=blue, urlcolor=blue, citecolor=blue}

\usepackage{caption} 
\usepackage{multicol} 

\begin{document}

\newcommand{\vsini}{$v\sin{i}$} 
\NewDocumentCommand{\vsin}{m}{%
  $v\sin{i} \!=\! #1\,\kmps$%
}
\newcommand{\kmps}{\si{\kilo\metre\per\second}} 
\newcommand{\mps}{\si{\metre\per\second}} 

\newcommand{\teff}{\mathrm{T$_{eff}$}} 
\newcommand{\logg}{\mathrm{$\log{g}$}} 

\lefttitle{S.R. Berlanas}
\righttitle{Massive stars in the era of large spectroscopic surveys}

\jnlPage{1}{7}
\jnlDoiYr{2021}
\doival{10.1017/xxxxx}

\aopheadtitle{Proceedings IAU Symposium}
\editors{A. Wofford, N. St-Louis, M. García \& S. Simón-Díaz, eds.}

\title{Massive stars in the era of large spectroscopic surveys: The MEIGAS project}

\author{S. R. Berlanas}
\affiliation{Centro de Astrobiología, CSIC-INTA. 28 692 Villanueva de la Cañada, Madrid, Spain}

\begin{abstract}
In the era of large spectroscopic surveys, a vast amount of spectra of massive stars will be gathered and supplemented by the wealth of astrometric and photometric data provided by the $Gaia$ satellite. Released data will mean a major step forward in the study of massive stars, giving us the chance to create statistically significant samples to explore the role of almost any parameter. In this contribution, I introduce to the community the \textit{ Multi-wavelength Exploration of massIve star-forminG regions and ASsociations} project (MEIGAS) and long-term plans for conducting comprehensive studies in the major galactic and near extragalactic star-forming regions and OB associations. Benefiting from current and forthcoming data from large scale spectroscopic surveys such as WEAVE and 4MOST (among others), as well as complementary observations at different wavelength ranges, the project aims to achieve crucial and complementary information to adequately characterize these regions and their stellar content, something imperative to improve our understanding of star formation and poorly known evolutionary pathways of massive stars.
\end{abstract}


\maketitle

\section{Introduction}
Hyper-energetic supernovae, Wolf-Rayet stars, red supergiants, massive stellar black holes,
neutron stars, long-duration gamma ray bursts, and the recently detected gravitational wave
sources are all direct descendants of massive OB stars \citep{woosley02}.  Despite their crucial role in driving the chemical and dynamical evolution of galaxies and the Universe, the evolutionary pathways of massive stars remain poorly known \citep{langer12}.\vspace{0.15cm}

To fully assess the impact of massive stars on the Universe, we must understand their physics and evolutionary behaviour. However,  large uncertainties still persist in our understanding of these objects that prevent us from fully understanding how massive stars impact other areas of Astrophysics. Key factors such as rotation, stellar winds and mass-loss dramatically affect their evolution. In addition to uncertainties related to individual models, massive stars are usually born in multiple systems \citep{sana12}, where the components are close enough to interact during their lifetime. As a result, the possible channels of evolution and final fate are multiplied \citep{demink13}, and their frequency needs to be determined. \vspace{0.15cm}

Star-forming regions represent ideal laboratories to study the processes of massive star
formation, evolution and their interaction with the interstellar medium (ISM) as they host large populations of OB stars observable across the entire electromagnetic spectrum that can be analyzed.
The availability of extensive multi-wavelength and multi-epoch observational datasets, combined with the analysis of large samples of high-quality OB-star spectra, offers a powerful means to constrain their physical properties and evolutionary states.\vspace{0.15cm}

Multi-wavelength observations are particularly valuable, as each spectral domain probes different physical processes and layers of the stellar atmosphere and its surroundings. By combining data from different wavelength ranges, we can obtain the most reliable constraints on key stellar parameters, such as effective temperature (T$_{eff}$), surface gravity ($\log{g}$), rotational velocity (\vsini), mass-loss rate ($\dot{M}$), wind clumping, helium abundance (He), and metallicity (Z), which are all key ingredients to unravel the evolutionary channels of massive stars.\vspace{0.15cm}

Within this context, the \textbf{Multi-wavelength Exploration of massIve star-forminG regions and Associations (MEIGAS) project}  aims at performing comprehensive studies in the major star-forming regions and OB associations through multi-wavelength observations.  Benefiting from current and forthcoming data from large scale spectroscopic surveys such as WEAVE \citep{jin24}, 4MOST \citep{dejong11}, GES \citep{randich22, gilmore22}, GOSSS \citep{maiz10} or IACOB \citep{ssimon11}, as well as complementary observations at different wavelength ranges \citep[e.g., from XShootU and BLOeM, see][]{vink23, shenar24}, the project aims to achieve crucial and complementary information to adequately characterize these regions and their stellar content, something imperative to improve our understanding of star formation and poorly known evolutionary pathways of massive stars.

\section{MEIGAS I: The universality of the vsini distribution}

The first phase of MEIGAS is focused on determining, for the first time ever, whether the empirical rotational distribution of massive O-type stars is universal -- as theory predicts \citep[see][]{demink13, demink14} -- or not, as suggested by recent findings in two of the main Galactic young OB associations, Cygnus OB2 and Carina OB1 \citep[see][]{berlanas20, berlanas25}. MEIGAS will inspect rotational properties of their massive stellar populations in different environments, providing the key chance to estimate the frequencies associated to the different evolutionary channels. These frequencies are written in the rotational distribution and in the stellar properties and abundances, and
although the census of possible evolutionary scenarios is not yet complete, we
aim for a first empirical estimation. \vspace{0.15cm}

\begin{figure}[h!]
        \centering
        \resizebox{0.65\hsize}{!}{\includegraphics[scale=1]{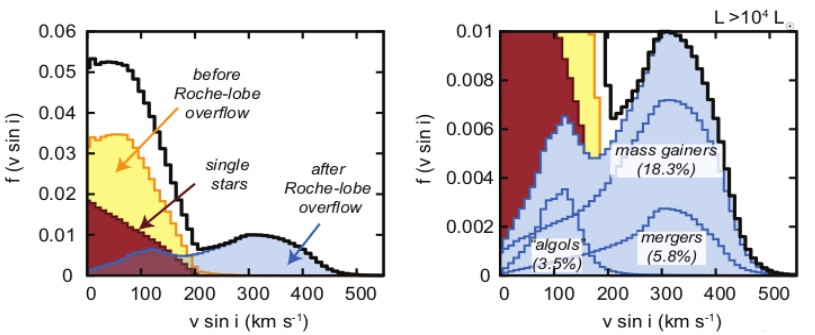}}
        \resizebox{0.45\hsize}{!}{\includegraphics[scale=1]{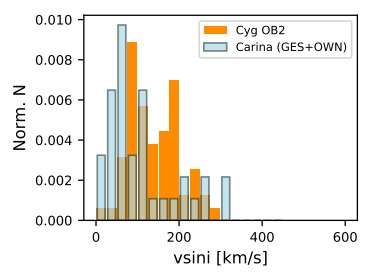}}
        \caption{From \cite{demink13} (top panels) and \cite{berlanas25} (bottom panel). Projected rotation rate (\vsini) distribution for massive O stars assuming continuous star formation. Top-right panel shows the cumulative distribution function indicating the possible evolutionary scenarios related to current rotational velocity. Top-left panel shows the full distribution function indicating the expected frequency of post-binary interaction. Bottom panel shows the empirical distribution for the O population in Cygnus OB2 and Carina OB1, evidencing a clear lack of fast rotators.
            }
        \label{vsini}
    \end{figure}

In recent years a considerable effort has been invested in empirically establishing the full distribution of rotational velocities of massive OB stars in different environments. Studies conducted in large regions such as the Milky Way \citep{ssimon14b, holgado22} and the 30 Doradus region in the Large Magellanic Cloud \citep{ragudelo13} have reported a similar bimodal distribution, consistent with the theoretical predictions of \cite{demink13, demink14} (see top panels of Fig.~\ref{vsini}). However, similar spectroscopic studies in young OB associations  such as Cygnus OB2 \citep[within the Cygnus-X complex, see ][]{berlanas20} and Car OB1 \citep[within the Carina Nebula, see][]{berlanas25} found a  significant  difference: a lack of fast rotating O stars (see bottom panel of Fig.~\ref{vsini}), a fact that poses a serious problem for present-day evolutionary models and scenarios of massive stars.\vspace{0.15cm}

MEIGAS will exploit already available data from, e.g., IACOB, GES, GOSSS, XShootU, BLOeM and own multi-wavelength spectra, together with the imminent arrival of high-quality spectra from large spectroscopic surveys such as WEAVE and 4MOST. Complementarily, current and forthcoming $Gaia$ data releases will provide crucial astrometric and photometric information. This combined multi-survey approach will enable the extension of the work we conducted in Cygnus OB2 and Carina OB1 \citep[see][]{berlanas18a, berlanas18b, berlanas19, berlanas20, herrero22, berlanas23, berlanas25}  to
the whole Cygnus-X and Carina complexes, as well as a detailed exploration of other major Galactic and near extragalactic star-forming regions at different wavelength ranges. This way, it will be possible to verify whether the significant differences found in both regions are due to specific environment properties related to young OB associations, such as incompleteness, extinction, or stellar ejections. The adding of two further extragalactic environments (LMC, SMC) will serve to address possible metallicity effect on the distribution.\vspace{0.9cm}

Main objectives of \textit{MEIGAS~I: The universality of the vsini distribution} are:
\begin{enumerate}
    \item  Inspect whether the significant differences found in Cygnus OB2 and Carina OB1 arise from specific environmental properties associated with young OB associations. Since completeness on the stellar census is crucial for this study, infrared observations and the identification of possible runaway stars ejected at high velocity from the core of stellar associations must be carefully addressed.\vspace{0.1cm}
    
    \item Assess whether the same deficiency of fast-rotating stars is found in other young star-forming regions, both within our Galaxy and beyond. Benefiting from already available high-quality spectroscopic data from surveys focused on massive stars, together with the imminent arrival of a huge amount of high-quality spectra from  WEAVE and 4MOST, we are expanding research to the whole complexes, Cygnus-X and the Carina Nebula. Furthermore, we will be able to extend research to other Galactic massive associations, such as NGC 6611, NGC 3603, Westerlund 1 and 2 or the Vela, Orion and Monoceros regions. The adding two further near extragalactic environments (LMC and SMC) to the study that will serve to address possible metallicity effect on the universality of the distribution of rotational velocities.
  
\end{enumerate}

\section{Current status}

We are conducting infrared observations in Cygnus OB2 and Carina OB1 using the EMIR and KMOS instruments at the GTC and VLT, respectively, to uncover the heavily obscured young massive population in these regions. Moreover, a detailed spectroscopic analysis is underway for a high-confidence sample of runaway candidates around Cygnus OB2, selected based on precise $Gaia$ DR3 astrometry and complementary observations. Notably, both the obscured members and the confirmed runaway populations in these associations may account for the missing population of fast rotators. In that case, we would recover the predicted rotational velocity distribution, suggesting that the frequencies of the possible evolutionary pathways are intrinsically similar but modulated by environmental conditions.\vspace{0.15cm}

In parallel, WEAVE, the new multi-object spectrograph (MOS, which is currently in its commissioning phase) for the 4.2-m William Herschel Telescope in la Palma, includes the Stellar, Circumstellar and Interstellar Physics (SCIP) survey, which will exploit its MOS capabilities to obtain mid-resolution spectra for thousands of stars along the northern Galactic disc. In addition, SCIP incorporates a dedicated high-resolution survey in the Cygnus-X complex, which will provide hundreds of multi-epoch spectra across the region. Complementarily, 4MOST, the new fibre-fed spectroscopic facility on the VISTA telescope, achieved its first light just a few weeks ago and will soon begin large-scale surveys of the southern sky, including key massive star-forming regions such as the Carina Nebula. Our membership in both collaborations will ensure early access to the datasets, allowing us to promptly integrate and analyse the forthcoming observations within the framework of the MEIGAS project.

\section{Summary}

MEIGAS will not only provide new empirical constraints on massive star evolution but will produce an impact across many fields of Astrophysics, from star and planet formation to galaxy and cluster evolution. This project will assess for the first time ever the universality of the rotational distribution of massive O stars and establish the relative frequency of the different channels of binary evolution under well known, homogeneous conditions, thus resolving a serious problem for present-day evolutionary models and scenarios of massive stars. Moreover, by establishing whether the distribution is environment-dependent or an universal function, this project will provide the necessary constraints for next-generation population synthesis models that are fundamental to simulating galaxy formation and evolution across cosmic time. 

{\small
\bibliographystyle{aa}   
\bibliography{iaus402}

\begin{thebibliography}{23}
\expandafter\ifx\csname natexlab\endcsname\relax\def\natexlab#1{#1}\fi

\bibitem[{{Berlanas} {et~al.}(2018{\natexlab{a}}){Berlanas}, {Herrero}, {Comer{\'o}n}, {Pasquali}, {Bertelli Motta}, \& {Sota}}]{berlanas18a}
{Berlanas}, S.~R., {Herrero}, A., {Comer{\'o}n}, F., {et~al.} 2018{\natexlab{a}}, \aap, 612, A50

\bibitem[{{Berlanas} {et~al.}(2018{\natexlab{b}}){Berlanas}, {Herrero}, {Comer{\'o}n}, {Sim{\'o}n-D{\'{\i}}az}, {Cervi{\~n}o}, \& {Pasquali}}]{berlanas18b}
{Berlanas}, S.~R., {Herrero}, A., {Comer{\'o}n}, F., {et~al.} 2018{\natexlab{b}}, \aap, 620, A56

\bibitem[{{Berlanas} {et~al.}(2020){Berlanas}, {Herrero}, {Comer{\'o}n}, {Sim{\'o}n-D{\'\i}az}, {Lennon}, {Pasquali}, {Ma{\'\i}z Apell{\'a}niz}, {Sota}, \& {Peller{\'\i}n}}]{berlanas20}
{Berlanas}, S.~R., {Herrero}, A., {Comer{\'o}n}, F., {et~al.} 2020, \aap, 642, A168

\bibitem[{{Berlanas} {et~al.}(2025){Berlanas}, {Mahy}, {Herrero}, {Ma{\'\i}z Apell{\'a}niz}, {Blomme}, {Comer{\'o}n}, {Negueruela}, {Molina Lera}, {Pantaleoni Gonz{\'a}lez}, {Daflon}, {Santos}, \& {Kalari}}]{berlanas25}
{Berlanas}, S.~R., {Mahy}, L., {Herrero}, A., {et~al.} 2025, \aap, 695, A248

\bibitem[{{Berlanas} {et~al.}(2023){Berlanas}, {Ma{\'\i}z Apell{\'a}niz}, {Herrero}, {Mahy}, {Blomme}, {Negueruela}, {Dorda}, {Comer{\'o}n}, {Gosset}, {Pantaleoni Gonz{\'a}lez}, {Molina Lera}, {Sota}, {Furst}, {Alfaro}, {Bergemann}, {Carraro}, {Drew}, {Morbidelli}, \& {Vink}}]{berlanas23}
{Berlanas}, S.~R., {Ma{\'\i}z Apell{\'a}niz}, J., {Herrero}, A., {et~al.} 2023, \aap, 671, A20

\bibitem[{{Berlanas} {et~al.}(2019){Berlanas}, {Wright}, {Herrero}, {Drew}, \& {Lennon}}]{berlanas19}
{Berlanas}, S.~R., {Wright}, N.~J., {Herrero}, A., {Drew}, J.~E., \& {Lennon}, D.~J. 2019, \mnras, 484, 1838

\bibitem[{{de Jong}(2011)}]{dejong11}
{de Jong}, R. 2011, The Messenger, 145, 14

\bibitem[{{de Mink} {et~al.}(2013){de Mink}, {Langer}, {Izzard}, {Sana}, \& {de Koter}}]{demink13}
{de Mink}, S.~E., {Langer}, N., {Izzard}, R.~G., {Sana}, H., \& {de Koter}, A. 2013, \apj, 764, 166

\bibitem[{{de Mink} {et~al.}(2014){de Mink}, {Sana}, {Langer}, {Izzard}, \& {Schneider}}]{demink14}
{de Mink}, S.~E., {Sana}, H., {Langer}, N., {Izzard}, R.~G., \& {Schneider}, F.~R.~N. 2014, \apj, 782, 7

\bibitem[{{Gilmore} {et~al.}(2022){Gilmore}, {Randich}, {Worley}, {Hourihane}, {Gonneau}, {Sacco}, {Lewis}, {Magrini}, {Fran{\c{c}}ois}, {Jeffries}, {Koposov}, {Bragaglia}, {Alfaro}, {Allende Prieto}, {Blomme}, {Korn}, {Lanzafame}, {Pancino}, {Recio-Blanco}, {Smiljanic}, {Van Eck}, {Zwitter}, {Bensby}, {Flaccomio}, {Irwin}, {Franciosini}, {Morbidelli}, {Damiani}, {Bonito}, {Friel}, {Vink}, {Prisinzano}, {Abbas}, {Hatzidimitriou}, {Held}, {Jordi}, {Paunzen}, {Spagna}, {Jackson}, {Ma{\'\i}z Apell{\'a}niz}, {Asplund}, {Bonifacio}, {Feltzing}, {Binney}, {Drew}, {Ferguson}, {Micela}, {Negueruela}, {Prusti}, {Rix}, {Vallenari}, {Bergemann}, {Casey}, {de Laverny}, {Frasca}, {Hill}, {Lind}, {Sbordone}, {Sousa}, {Adibekyan}, {Caffau}, {Daflon}, {Feuillet}, {Gebran}, {Gonzalez Hernandez}, {Guiglion}, {Herrero}, {Lobel}, {Merle}, {Mikolaitis}, {Montes}, {Morel}, {Ruchti}, {Soubiran}, {Tabernero}, {Tautvai{\v{s}}ien{\.{e}}}, {Traven}, {Valentini}, {Van der Swaelmen}, {Villanova}, {Viscasillas V{\'a}zquez}, {Bayo},
  {Biazzo}, {Carraro}, {Edvardsson}, {Heiter}, {Jofr{\'e}}, {Marconi}, {Martayan}, {Masseron}, {Monaco}, {Walton}, {Zaggia}, {Aguirre B{\o}rsen-Koch}, {Alves}, {Balaguer-Nunez}, {Barklem}, {Barrado}, {Bellazzini}, {Berlanas}, {Binks}, {Bressan}, {Capuzzo-Dolcetta}, {Casagrande}, {Casamiquela}, {Collins}, {D'Orazi}, {Dantas}, {Debattista}, {Delgado-Mena}, {Di Marcantonio}, {Drazdauskas}, {Evans}, {Famaey}, {Franchini}, {Fr{\'e}mat}, {Fu}, {Geisler}, {Gerhard}, {Gonz{\'a}lez Solares}, {Grebel}, {Guti{\'e}rrez Albarr{\'a}n}, {Jim{\'e}nez-Esteban}, {J{\"o}nsson}, {Khachaturyants}, {Kordopatis}, {Kos}, {Lagarde}, {Ludwig}, {Mahy}, {Mapelli}, {Marfil}, {Martell}, {Messina}, {Miglio}, {Minchev}, {Moitinho}, {Montalban}, {Monteiro}, {Morossi}, {Mowlavi}, {Mucciarelli}, {Murphy}, {Nardetto}, {Ortolani}, {Paletou}, {Palou{\v{s}}}, {Pickering}, {Quirrenbach}, {Re Fiorentin}, {Read}, {Romano}, {Ryde}, {Sanna}, {Santos}, {Seabroke}, {Spina}, {Steinmetz}, {Stonkut{\'e}}, {Sutorius}, {Th{\'e}venin}, {Tosi}, {Tsantaki},
  {Wright}, {Wyse}, {Zoccali}, {Zorec}, \& {Zucker}}]{gilmore22}
{Gilmore}, G., {Randich}, S., {Worley}, C.~C., {et~al.} 2022, \aap, 666, A120

\bibitem[{{Herrero} {et~al.}(2022){Herrero}, {Berlanas}, {Gil de Paz}, {Comer{\'o}n}, {Puls}, {Ram{\'\i}rez Alegr{\'\i}a}, {Garcia}, {Lennon}, {Najarro}, {Sim{\'o}n-D{\'\i}az}, {Urbaneja}, {Gallego}, {Carrasco}, {Iglesias}, {Cedazo}, {Garc{\'\i}a Vargas}, {Castillo-Morales}, {Pascual}, {Cardiel}, {P{\'e}rez-Calpena}, {G{\'o}mez-Alvarez}, \& {Mart{\'\i}nez-Delgado}}]{herrero22}
{Herrero}, A., {Berlanas}, S.~R., {Gil de Paz}, A., {et~al.} 2022, \mnras, 511, 3113

\bibitem[{{Holgado} {et~al.}(2022){Holgado}, {Sim{\'o}n-D{\'\i}az}, {Herrero}, \& {Barb{\'a}}}]{holgado22}
{Holgado}, G., {Sim{\'o}n-D{\'\i}az}, S., {Herrero}, A., \& {Barb{\'a}}, R.~H. 2022, \aap, 665, A150

\bibitem[{{Jin} {et~al.}(2024){Jin}, {Trager}, {Dalton}, {Aguerri}, {Drew}, {Falc{\'o}n-Barroso}, {G{\"a}nsicke}, {Hill}, {Iovino}, {Pieri}, {Poggianti}, {Smith}, {Vallenari}, {Abrams}, {Aguado}, {Antoja}, {Arag{\'o}n-Salamanca}, {Ascasibar}, {Babusiaux}, {Balcells}, {Barrena}, {Battaglia}, {Belokurov}, {Bensby}, {Bonifacio}, {Bragaglia}, {Carrasco}, {Carrera}, {Cornwell}, {Dom{\'\i}nguez-Palmero}, {Duncan}, {Famaey}, {Fari{\~n}a}, {Gonzalez}, {Guest}, {Hatch}, {Hess}, {Hoskin}, {Irwin}, {Knapen}, {Koposov}, {Kuchner}, {Laigle}, {Lewis}, {Longhetti}, {Lucatello}, {M{\'e}ndez-Abreu}, {Mercurio}, {Molaeinezhad}, {Mongui{\'o}}, {Morrison}, {Murphy}, {Peralta de Arriba}, {P{\'e}rez}, {P{\'e}rez-R{\`a}fols}, {Pic{\'o}}, {Raddi}, {Romero-G{\'o}mez}, {Royer}, {Siebert}, {Seabroke}, {Som}, {Terrett}, {Thomas}, {Wesson}, {Worley}, {Alfaro}, {Allende Prieto}, {Alonso-Santiago}, {Amos}, {Ashley}, {Balaguer-N{\'u}{\~n}ez}, {Balbinot}, {Bellazzini}, {Benn}, {Berlanas}, {Bernard}, {Best}, {Bettoni}, {Bianco}, {Bishop},
  {Blomqvist}, {Boeche}, {Bolzonella}, {Bonoli}, {Bosma}, {Britavskiy}, {Busarello}, {Caffau}, {Cantat-Gaudin}, {Castro-Ginard}, {Couto}, {Carbajo-Hijarrubia}, {Carter}, {Casamiquela}, {Conrado}, {Corcho-Caballero}, {Costantin}, {Deason}, {de Burgos}, {De Grandi}, {Di Matteo}, {Dom{\'\i}nguez-G{\'o}mez}, {Dorda}, {Drake}, {Dutta}, {Erkal}, {Feltzing}, {Ferr{\'e}-Mateu}, {Feuillet}, {Figueras}, {Fossati}, {Franciosini}, {Frasca}, {Fumagalli}, {Gallazzi}, {Garc{\'\i}a-Benito}, {Gentile Fusillo}, {Gebran}, {Gilbert}, {Gledhill}, {Gonz{\'a}lez Delgado}, {Greimel}, {Guarcello}, {Guerra}, {Gullieuszik}, {Haines}, {Hardcastle}, {Harris}, {Haywood}, {Helmi}, {Hernandez}, {Herrero}, {Hughes}, {Ir{\v{s}}i{\v{c}}}, {Jablonka}, {Jarvis}, {Jordi}, {Kondapally}, {Kordopatis}, {Krogager}, {La Barbera}, {Lam}, {Larsen}, {Lemasle}, {Lewis}, {Lhom{\'e}}, {Lind}, {Lodi}, {Longobardi}, {Lonoce}, {Magrini}, {Ma{\'\i}z Apell{\'a}niz}, {Marchal}, {Marco}, {Martin}, {Matsuno}, {Maurogordato}, {Merluzzi}, {Miralda-Escud{\'e}},
  {Molinari}, {Monari}, {Morelli}, {Mottram}, {Naylor}, {Negueruela}, {O{\~n}orbe}, {Pancino}, {Peirani}, {Peletier}, {Pozzetti}, {Rainer}, {Ramos}, {Read}, {Rossi}, {R{\"o}ttgering}, {Rubi{\~n}o-Mart{\'\i}n}, {Sabater}, {San Juan}, {Sanna}, {Schallig}, {Schiavon}, {Schultheis}, {Serra}, {Shimwell}, {Sim{\'o}n-D{\'\i}az}, {Smith}, {Sordo}, {Sorini}, {Soubiran}, {Starkenburg}, {Steele}, {Stott}, {Stuik}, {Tolstoy}, {Tortora}, {Tsantaki}, {Van der Swaelmen}, {van Weeren}, \& {Vergani}}]{jin24}
{Jin}, S., {Trager}, S.~C., {Dalton}, G.~B., {et~al.} 2024, \mnras, 530, 2688

\bibitem[{{Langer}(2012)}]{langer12}
{Langer}, N. 2012, \araa, 50, 107

\bibitem[{{Ma{\'{\i}}z Apell{\'a}niz}(2010)}]{maiz10}
{Ma{\'{\i}}z Apell{\'a}niz}, J. 2010, \aap, 518, A1

\bibitem[{{Ram{\'{\i}}rez-Agudelo} {et~al.}(2013){Ram{\'{\i}}rez-Agudelo}, {Sim{\'o}n-D{\'{\i}}az}, {Sana}, {de Koter}, {Sab{\'{\i}}n-Sanjul{\'{\i}}an}, {de Mink}, {Dufton}, {Gr{\"a}fener}, {Evans}, {Herrero}, {Langer}, {Lennon}, {Ma{\'{\i}}z Apell{\'a}niz}, {Markova}, {Najarro}, {Puls}, {Taylor}, \& {Vink}}]{ragudelo13}
{Ram{\'{\i}}rez-Agudelo}, O.~H., {Sim{\'o}n-D{\'{\i}}az}, S., {Sana}, H., {et~al.} 2013, \aap, 560, A29

\bibitem[{{Randich} {et~al.}(2022){Randich}, {Gilmore}, {Magrini}, {Sacco}, {Jackson}, {Jeffries}, {Worley}, {Hourihane}, {Gonneau}, {Viscasillas Vazquez}, {Franciosini}, {Lewis}, {Alfaro}, {Allende Prieto}, {Bensby}, {Blomme}, {Bragaglia}, {Flaccomio}, {Fran{\c{c}}ois}, {Irwin}, {Koposov}, {Korn}, {Lanzafame}, {Pancino}, {Recio-Blanco}, {Smiljanic}, {Van Eck}, {Zwitter}, {Asplund}, {Bonifacio}, {Feltzing}, {Binney}, {Drew}, {Ferguson}, {Micela}, {Negueruela}, {Prusti}, {Rix}, {Vallenari}, {Bayo}, {Bergemann}, {Biazzo}, {Carraro}, {Casey}, {Damiani}, {Frasca}, {Heiter}, {Hill}, {Jofr{\'e}}, {de Laverny}, {Lind}, {Marconi}, {Martayan}, {Masseron}, {Monaco}, {Morbidelli}, {Prisinzano}, {Sbordone}, {Sousa}, {Zaggia}, {Adibekyan}, {Bonito}, {Caffau}, {Daflon}, {Feuillet}, {Gebran}, {Gonzalez Hernandez}, {Guiglion}, {Herrero}, {Lobel}, {Maiz Apellaniz}, {Merle}, {Mikolaitis}, {Montes}, {Morel}, {Soubiran}, {Spina}, {Tabernero}, {Tautvai{\v{s}}iene}, {Traven}, {Valentini}, {Van der Swaelmen}, {Villanova}, {Wright},
  {Abbas}, {Aguirre B{\o}rsen-Koch}, {Alves}, {Balaguer-Nunez}, {Barklem}, {Barrado}, {Berlanas}, {Binks}, {Bressan}, {Capuzzo-Dolcetta}, {Casagrande}, {Casamiquela}, {Collins}, {D'Orazi}, {Dantas}, {Debattista}, {Delgado-Mena}, {Di Marcantonio}, {Drazdauskas}, {Evans}, {Famaey}, {Franchini}, {Fr{\'e}mat}, {Friel}, {Fu}, {Geisler}, {Gerhard}, {Gonzalez Solares}, {Grebel}, {Gutierrez Albarran}, {Hatzidimitriou}, {Held}, {Jim{\'e}nez-Esteban}, {J{\"o}nsson}, {Jordi}, {Khachaturyants}, {Kordopatis}, {Kos}, {Lagarde}, {Mahy}, {Mapelli}, {Marfil}, {Martell}, {Messina}, {Miglio}, {Minchev}, {Moitinho}, {Montalban}, {Monteiro}, {Morossi}, {Mowlavi}, {Mucciarelli}, {Murphy}, {Nardetto}, {Ortolani}, {Paletou}, {Palou{\v{s}}}, {Paunzen}, {Pickering}, {Quirrenbach}, {Re Fiorentin}, {Read}, {Romano}, {Ryde}, {Sanna}, {Santos}, {Seabroke}, {Spagna}, {Steinmetz}, {Stonkut{\'e}}, {Sutorius}, {Th{\'e}venin}, {Tosi}, {Tsantaki}, {Vink}, {Wright}, {Wyse}, {Zoccali}, {Zorec}, {Zucker}, \& {Walton}}]{randich22}
{Randich}, S., {Gilmore}, G., {Magrini}, L., {et~al.} 2022, \aap, 666, A121

\bibitem[{{Sana} {et~al.}(2012){Sana}, {de Mink}, {de Koter}, {Langer}, {Evans}, {Gieles}, {Gosset}, {Izzard}, {Le Bouquin}, \& {Schneider}}]{sana12}
{Sana}, H., {de Mink}, S.~E., {de Koter}, A., {et~al.} 2012, Science, 337, 444

\bibitem[{{Shenar} {et~al.}(2024){Shenar}, {Bodensteiner}, {Sana}, {Crowther}, {Lennon}, {Abdul-Masih}, {Almeida}, {Backs}, {Berlanas}, {Bernini-Peron}, {Bestenlehner}, {Bowman}, {Bronner}, {Britavskiy}, {de Koter}, {de Mink}, {Deshmukh}, {Evans}, {Fabry}, {Gieles}, {Gilkis}, {Gonz{\'a}lez-Tor{\`a}}, {Gr{\"a}fener}, {G{\"o}tberg}, {Hawcroft}, {H{\'e}nault-Brunet}, {Herrero}, {Holgado}, {Janssens}, {Johnston}, {Josiek}, {Justham}, {Kalari}, {Katabi}, {Keszthelyi}, {Klencki}, {Kub{\'a}t}, {Kub{\'a}tov{\'a}}, {Langer}, {Lefever}, {Ludwig}, {Mackey}, {Mahy}, {Ma{\'\i}z Apell{\'a}niz}, {Mandel}, {Maravelias}, {Marchant}, {Menon}, {Najarro}, {Oskinova}, {O'Grady}, {Ovadia}, {Patrick}, {Pauli}, {Pawlak}, {Ramachandran}, {Renzo}, {Rocha}, {Sander}, {Sayada}, {Schneider}, {Schootemeijer}, {Sch{\"o}sser}, {Sch{\"u}rmann}, {Sen}, {Shahaf}, {Sim{\'o}n-D{\'\i}az}, {Stoop}, {Toonen}, {Tramper}, {van Loon}, {Valli}, {van Son}, {Vigna-G{\'o}mez}, {Villase{\~n}or}, {Vink}, {Wang}, \& {Willcox}}]{shenar24}
{Shenar}, T., {Bodensteiner}, J., {Sana}, H., {et~al.} 2024, \aap, 690, A289

\bibitem[{{Sim{\'o}n-D{\'{\i}}az} {et~al.}(2011){Sim{\'o}n-D{\'{\i}}az}, {Castro}, {Herrero}, {Puls}, {Garcia}, \& {Sab{\'{\i}}n-Sanjuli{\'a}n}}]{ssimon11}
{Sim{\'o}n-D{\'{\i}}az}, S., {Castro}, N., {Herrero}, A., {et~al.} 2011, in Journal of Physics Conference Series, Vol. 328, Journal of Physics Conference Series, 012021

\bibitem[{{Sim{\'o}n-D{\'{\i}}az} {et~al.}(2014){Sim{\'o}n-D{\'{\i}}az}, {Herrero}, {Sab{\'{\i}}n-Sanjuli{\'a}n}, {Najarro}, {Garcia}, {Puls}, {Castro}, \& {Evans}}]{ssimon14b}
{Sim{\'o}n-D{\'{\i}}az}, S., {Herrero}, A., {Sab{\'{\i}}n-Sanjuli{\'a}n}, C., {et~al.} 2014, \aap, 570, L6

\bibitem[{{Vink} {et~al.}(2023){Vink}, {Mehner}, {Crowther}, {Fullerton}, {Garcia}, {Martins}, {Morrell}, {Oskinova}, {St-Louis}, {ud-Doula}, {Sander}, {Sana}, {Bouret}, {Kub{\'a}tov{\'a}}, {Marchant}, {Martins}, {Wofford}, {van Loon}, {Grace Telford}, {G{\"o}tberg}, {Bowman}, {Erba}, {Kalari}, {Abdul-Masih}, {Alkousa}, {Backs}, {Barbosa}, {Berlanas}, {Bernini-Peron}, {Bestenlehner}, {Blomme}, {Bodensteiner}, {Brands}, {Evans}, {David-Uraz}, {Driessen}, {Dsilva}, {Geen}, {G{\'o}mez-Gonz{\'a}lez}, {Grassitelli}, {Hamann}, {Hawcroft}, {Herrero}, {Higgins}, {John Hillier}, {Ignace}, {Istrate}, {Kaper}, {Kee}, {Kehrig}, {Keszthelyi}, {Klencki}, {de Koter}, {Kuiper}, {Laplace}, {Larkin}, {Lefever}, {Leitherer}, {Lennon}, {Mahy}, {Ma{\'\i}z Apell{\'a}niz}, {Maravelias}, {Marcolino}, {McLeod}, {de Mink}, {Najarro}, {Oey}, {Parsons}, {Pauli}, {Pedersen}, {Prinja}, {Ramachandran}, {Ram{\'\i}rez-Tannus}, {Sabhahit}, {Schootemeijer}, {Reyero Serantes}, {Shenar}, {Stringfellow}, {Sudnik}, {Tramper}, \& {Wang}}]{vink23}
{Vink}, J.~S., {Mehner}, A., {Crowther}, P.~A., {et~al.} 2023, \aap, 675, A154

\bibitem[{{Woosley} {et~al.}(2002){Woosley}, {Heger}, \& {Weaver}}]{woosley02}
{Woosley}, S.~E., {Heger}, A., \& {Weaver}, T.~A. 2002, Reviews of Modern Physics, 74, 1015

\end{thebibliography}
}

\end{document}